\documentclass[english,11pt]{article}

\usepackage{amssymb}
\usepackage{graphicx}
\usepackage{amsfonts}
\usepackage{bbm}

\usepackage{babel}
\usepackage{color}

\newcommand{\be}{\begin{equation}}
\newcommand{\ee}{\end{equation}}
\newcommand{\rf}[1]{(\ref{eq:#1})}
\newcommand{\X}{{\cal X}}

\newcommand{\G}{{\cal G}}

\begin{document}

\title{Applications of Nash's  Theorem to Cosmology\vspace{4mm}\\{\small Talk presented  at the  II  Indo-Brazilian  Workshop  on  Gravitation  and  Cosmology,\\  Natal, Brazil, October  2008}}

\author{M. D. Maia\thanks{maia@unb.br} \\
Instituto  de F\'\i sica, Universidade de Bras\'\i lia,
  Bras\'\i lia, 70919-970, Brazil. }

\maketitle

\abstract{ The  cosmological  constant problem  is seen  as  a  symptom  of the  ambiguity  of  the  Riemann  curvature in general  relativity. The solution of that  ambiguity  provided  by Nash's  theorem on gravitational perturbations  along extra  dimensions  eliminate  the direct  comparison  between  the  vacuum  energy density and  Einstein's  cosmological  constant,  besides being  compatible  with  the formation of  structures  and  the  accelerated  expansion of  the universe.}

\tableofcontents

\section{The Gravitational Constant}
The  Gravitational  constant     in  Newton's  gravitational  law
 \be\vec{F} = -m a  =  -G  \frac{m  m'}{r^2}  \frac{\vec{r}}{r}
\label{eq:Newton}
\ee
was  introduced  to   convert  the physical  dimensions  $[M^2]/[L^2]$ to the  dimensions of  force $[M][L]/[T^2]$.
 It has  the value  $ G =6.67 \times 10^{-8}  cm^3/ g sec^2$,  with
   the same  value  in a wide range of applications of  \rf{Newton}.
In  1914  Max  Planck suggested  a  natural  units  system  in which   $G=c=\hbar=1$  and  everything  else would be  measured in centimeters. For that purpose it  was  assumed  that Newton's  equation  \rf{Newton} also  holds   at the quantum level. Under this  condition, comparing the   gravitational energy for $  m=m'$ with the  quantum of energy for  a  wavelength   $ \lambda  \approx r $, it follows  that
$$
E=  <\vec{F}.\vec{r} > =  G  \frac{m^2}{\lambda}= \frac{\hbar c}{\lambda}
$$
Together  with Maxwell  equations  and  the  laws of  thermodynamics,   this  leads  to three  quantities  which characterize the  so called Planck regime:
\be
m_{P\ell} =  \sqrt{\frac{\hbar  c}{G}}\approx  10^{19} Gev,\;\;\;
\lambda_{P\ell} = \sqrt{\frac{\hbar G }{c^3}} \approx 10^{-33}cm,\;\;\;
t_{P\ell}  = \sqrt{\frac{\hbar G }{c^5}} \approx  10^{-44}sec  \label{eq:Planck}
\ee
Planck's  conclusion    established  a  landmark in the development of   modern physics:\\
"\emph{These quantities retain their natural significance as long as the  law of gravitation and that of the propagation of light in a vacuum and the two principles of thermodynamics remain valid; they therefore must be found always the same, when measured by the most widely differing intelligences according to the most widely differing methods}"  \cite{Planck}.

Today,  we can safely  say  that  electrodynamics, actually all known gauge theories, and  the laws  of  thermodynamics  remain solid.  However, the  validity of Newton's  law   at   $10^{-33}$cm  has  not been experimentally  confirmed.  It has been  recently shown  to  hold  at     $10^{-3}$cm,  but with   strong hints   that it  breaks  down  at   $10^{-4}$cm   \cite{Decca}. It  should be noted  also that the  constant    $G$  is  valid   for the Newtonian  space-time  which has   the product  topology  $\Sigma_3 \times I\!\!R$,  where  $\Sigma_3$  denotes the      3-dimensional  simultaneity  sections,  implying  that  the  gravitational  constant  has the physical dimensions  $[G]= [L]^3/[M][T]^2$,    appropriate  for  3-dimensional  manifolds  only.

  In 1916 Newton's  gravitational  law  changed  dramatically  to general relativity, including  the principles  of  equivalence,  the  general  covariance and
Einstein's  equations in  a  4-dimensional  space-time
\be
 R_{\mu\nu}-\frac{1}{2}R g_{\mu\nu}  =8\pi G  T_{\mu\nu}  \label{eq:Einstein}
\ee
The   Newtonian gravitational  constant   $G$,   was  retained  in  \rf{Einstein}, to guarantee   that  the  theory  would  reproduce  the Newtonian theory in its   weak  field  limit,  without the  need  to change  constants. However, the  consequences of this  are  quite  embarrassing:  Indeed,  the   maintenance of $G$  in   \rf{Einstein}  originates  the hierarchy problem of the fundamental  interactions. While  all relativistic  gauge   interactions   are  quantized
 at the   Tev   scale of  energies,  gravitation  would be  quantized  only  at  $\approx  10^{19}Gev$,  which, as  we have seen,   coincide with the  energy level predicted by Planck  for Newtonian quantum gravity, which is the weak  field  limit of general relativity. Furthermore,  the  relativistic    quantum  gravitational theory compatible  with  the  physical  dimensions of  $G$ would  be  defined only  in   a  3-dimensional foliation  of   space-time,  as  originally conceived by   Dirac  \cite{Dirac},    Arnowitt, Deser  and  Misner  \cite{ADM}. However,  such  foliation is  not  consistent   with  the  diffeomorphism invariance  of  general relativity \cite{Kuchar}.

The  criticism   on the validity of   Planck's  regime   for  quantum gravity  is  basis  of the seminal paper    by   Nima Arkani-Hamed,    Gia  Dvali \& Savas  Dimopolous (ADD) \cite{ADD}  in  1998,   inaugurating  the    brane-world  program:  Using   the known fact that all  gauge fields  are   defined in  four  dimensions, these  authors   proposed  a  solution of the  hierarchy problem,  assuming that  the four  dimensional  space-time  in  which  the known  gauge interactions  are  confined is    a  submanifold  embedded in a D=4+N-dimensional  space,  but gravitation   would  access  the extra dimensions,   with the  same  energy (Tev) of the  gauge interactions.   In the  following  we  show  how  Nash's  theorem  provides   a    mathematical  support  for  such program,  including  the  necessity  for  gravitational propagation in the  extra dimensions.

\section{Nash's  Theorem  and  Gravitational Perturbations}
As  far  as    the standard  gauge  theories are  concerned, the  4-dimensionality of  space-time  is  sufficient. This  can  be   understood in more general terms  as  a  consequence of the  duality   operations of the   Yang-Mills  equations. Defining  the   curvature 2-form
$F=F_{\mu\nu}dx^\mu \wedge dx^\nu,\;\;
F_{\mu\nu}=  [D_\mu,D_\nu],  \;\; D_\mu =\partial_\mu  + A_\mu$
then the Yang-Mills  equations are   written as
$$  D\wedge F=0\;\;  \mbox{e}\;\;    D\wedge F^*  =4\pi j^*$$
where   the  dual of the gauge  curvature   2-form  $F$  is
$$
F^*   =F^{*}_{\mu\nu} dx^\mu \wedge dx^\nu,\;\; F^*_{\mu\nu} =\epsilon_{\mu\nu\rho\sigma}F^{\rho\sigma}
$$
It follows that $D\wedge F^*$  is  a  3-form that  must be isomorphic to the current 1-form   $j^*$. This condition  depends  on the  4-dimensionality of the  space-time  \cite{Donaldson,Taubes}.
\vspace{2mm}\\
On the other hand,  the gravitational  field  not being  a standard gauge  theory does not have  such formal limitation.  Therefore,  it  was  assumed in  \cite{ADD} that  gravity  should propagate  along  extra  dimensions. This is  not al  all  obvious, but   it has
a  solid  mathematical justification  based in  the foundations of  Riemannian  geometry. The local curvature  of  a manifold  is given  by  the Riemann tensor
\be
{\cal R}(U,V)W = [\nabla_u,\nabla_V ]W - \nabla_{[U,V]}W \label{eq:Riemann}
\ee
which  describes the  variation of a  tangent vector  field  $W$  when it is  dragged  along a  four-sided  contour  defined  by  the vectors $U$  and  $V$. This   tensor  tells  about   the local  shape of the manifold,  without  comparing it  to anything  else that could be used as a  standard  of  shape.  This  implies that  \rf{Riemann} describe the  curvature of  a  class of  equivalence of  Riemannian  geometries with the same curvature, as was  stated by
Riemann  himself \cite{Riemann}:
"\emph{...arbitrary  cylindrical  or  conical  surfaces \emph{[manifolds]} count as  equivalent to a plane}..."
In other  words,  the  Riemann  curvature is  ambiguous in the sense that Riemannian manifolds   with  different local shapes  have the same  Riemannian  curvature.  This  situation  created  some discomfort  until   1873 when  L. Schlaefli  proposed  a solution for such  ambiguous  situation,  conjecturing that  all Riemannian manifold should  be embedded into a  larger (Euclidean) space, the latter acting as  the curvature reference \cite{Schlaefli}.

General relativity does not  present such   curvature  ambiguity  because  Minkowski's space-time was   set  as   the standard flat manifold against  which  all  other  space-time curvatures  are   compared. The same  space-time  is   chosen   as
 the   ground state for the gravitational field,  where    particles and  quantum  fields  are  defined.  This  choice  would  be  fine,  were  not  for  the experimental  evidences  of a small but  non-zero   cosmological  constant.  Since the presence of this  constant is  not    compatible  with  the   Minkowski  space-time,  we face   a  conflicting  situation: Either we  define   particles,   quantum  fields  and their  vacua states in the  Minkowski  space-time  using the  Poincaré group,   or  else these properties  should be defined   in  a  deSitter  space-time  using  the deSitter group \cite{MaiaFriedman}.  The  cosmological constant  and  the
vacuum energy density based on the Poincaré  symmetry  cannot be   present  simultaneously  in  Einstein's  equations, without  bringing up the   current   cosmological  constant  issue.

On the other hand,   the  embedding space suggested by Schlaefli can  provide a  curvature  reference independently of the ground state, thus eliminating the  conflict.  To see how this  works,  consider the   embedding map  ${\cal X}: V_n\rightarrow V_D$,   $D=n+N$ satisfying the   embedding  equations  (Index  convention:  $\mu, \nu  =  1..n,\; a,b = n+1..D,  \; A, B  = 1...D$)
$$
{\X}^{A}_{,\mu}{\X}^{B}_{,\nu}{\cal G}_{AB} ={g}_{\mu\nu},\;\; {\X}^{A}_{,\mu}{\eta}^{B}_{b}{\cal G}_{AB}=0, \;\;{\eta}^{A}_{a}{\eta}^{B}_{b}{\cal G}_{AB}={g}_{ab}
$$
were  $\eta_ a$ denotes   $N$   vectors orthogonal to  $V_n$.
When  the components of the Riemann tensor  of the  embedding space  ${\cal R}_{ABCD}$ are   written in the  Gaussian  frame $\{{\cal X}^A_{,\mu}, \eta^B_b \}$,  we obtain  the Gauss, Codazzi and Ricci equations  respectively
\begin{eqnarray}
&&{\cal R}_{ABCD}{\cal X}^{A}_{,\alpha}
{\cal X}^{B}_{,\beta}{\cal X}^{C}_{,\gamma}{\cal X}^{D}_{,\delta} =R_{\alpha\beta\gamma\delta} -
2g^{mn}k_{\alpha[\gamma m}k_{\delta]\beta n}  \label{eq:G}
\\
&&{\cal R}_{ABCD} {\cal
X}^{A}_{,\alpha} \eta^{B}_{b}{\cal X}^{C}_{,\gamma}{\cal
X}^{D}_{,\delta} =k_{\alpha[\gamma b; \delta]} -
g^{mn}A_{[\gamma mb}k_{\alpha\delta]n } \label{eq:C}\\
&&{\cal
R}_{ABCD}\eta^{A}_{a}\eta^{B}_{b} {\cal X}^{C}_{,\gamma} {\cal
X}^{D}_{,\delta} = -2g^{mn}A_{[\gamma ma}A_{\delta]n b} -2A_{[\gamma a b ; \delta]} - g^{mn}k_{[\gamma m a}k_{\delta]nb}  \label{eq:R}
\end{eqnarray}
showing the  components  of the  Riemann  curvature  of  $V_D$
expressed   in terms  of  the  Riemann tensor  of  $V_n$;  the
extrinsic  curvatures  $k_{\mu\nu a}$   (one for  each  $\eta_a$) and   the  components  of the third  fundamental  form $A_{\mu  ab}$.

However, if the   embedding space is  to  serve as  a reference
it must hold   for  all  possible Riemannian geometries.
This  relevant  detail  is  the main result of  Nash's theorem of  1956,  where  he  introduced  the concept of  metric perturbations  (or  deformations) along  the extra embedding dimensions  \cite{Nash}:  Given the     embedding
of some  background Riemannian geometry $V_n$    with metric $g_{\mu\nu}$, then  we    may  obtain another manifold  with the perturbed  metric    given by
\be
g_{\mu\nu}\rightarrow  g_{\mu\nu}  +  \delta g_{\mu\nu},\; \delta g_{\mu\nu} =-2k_{\mu\nu a} \delta y^a  \label{eq:pertu}
\ee
Thus,  Nash's perturbations   not  only warps    the  manifold, but it also stretches  deforms  it,   so that   all  Riemannian geometries  can be generated,  having the  curvature of the embedding space  as  a  reference.  Nash's theorem  was soon generalized to  pseudo-Riemannian  spaces embedded in pseudo-Riemannian Manifolds \cite{Greene}.

In order to   define  such  perturbations  the embedding must be
 differentiable. This condition was  implemented  by Nash,  using
an smoothing operation to eliminate edges  and  wrinkles  during the deformation.  We find  it  easier  and  perhaps    more  fundamental to  understand this  condition as   a  consequence of the   Einstein-Hilbert  principle  for the embedding space  geometry:
\be
\delta  A  =   \delta \int{{\cal  R} \sqrt{-{\cal G}}}dV  =0  \label{eq:EH}
\ee
which  has the meaning that   the embedding  space  has  the  smoothest   possible  curvature. Under such  condition  and
from  equations \rf{G}-\rf{R},   it  follows    that   the  embedding map  $\X$  must  also  be  differentiable to produce  a differentiable and  continuously perturbable embedded  manifold.
notice that  in order  to reproduce Einstein's gravitation, the    dynamical  variables  in   \rf{EH}  must be  taken  to be   the components of  the  embedding space  metric   $\G_{AB}$,  which  are  separated into   the variables $g_{\mu\nu}$,  $k_{\mu\nu a}$  and   $A_{\mu  ab}$  present  in  \rf{G}-\rf{R}.

\section{Cosmological Applications}

The   standard  Friedman-Lemaitre-Robertson-Walker (FLRW)  model is  sufficiently simple   to make it  locally  embedded in  a  5-dimensional  flat  space,  satisfying  Nash's  differentiable   conditions.  Therefore, it can be  taken  as  a  background cosmology, which  can be deformed   along the  fifth-dimension. However here  we  evaluate  the  effects  of  the extrinsic geometry in  the  FLWR  background  only (that is  without perturbations).
From   \rf{EH}  we obtain the 5-dimensional  Einstein's  equations \[
{\cal  R}_{AB}-\frac{1}{2}{\cal  R}{\cal  G}_{AB}  =8\pi G_* T^*_{AB}, \;\;\; \; A,B  =1..5
\]
where   $G_*$  denotes  the   gravitational  constant,  with  physical  dimensions appropriate  for  4-dimensional  hypersurfaces  of  the  5-dimensional space,    consistent  with  the  proposed  solution of the   hierarchy  problem.  Therefore,   the  value  of  $G$ is taken to  correspond  to  the  new Planck  scale, somewhere  within the  Tev  scale of  energies \cite{ADD}.
$T^*_{AB}$   denotes   the energy-momentum  tensor of    the    known  gauge  fields  and  of  ordinary matter,  both   confined to the  4-dimensional  space-time.  Since  only  gravitation  can  access   the extra  dimensions,  general  covariance   cannot apply  in  the  embedding space,   but  it   remains valid    in the  4-dimensional  space-time.
 When  the  above  equations are  written  in the  Gaussian  frame  $\{{\cal X}^A_{,\mu}, \eta^B \}$,   we  obtain  the 4-dimensional gravitational   equations  (for  the  derivation of these  equations  see  eg  \cite{GDEI}),
\begin{eqnarray}
&&R_{\mu\nu}-\frac{1}{2}Rg_{\mu\nu}+\Lambda g_{\mu\nu} +
{Q}_{\mu\nu} = 8\pi G_* T_{\mu\nu}\label{eq:gtensor}\\
&&k_{\mu ;\rho}^{\rho}\!  -\!h_{,\mu} = 0 \label{eq:gvector}
\end{eqnarray}
where we  have  simplified  the notation to $k_{\mu\nu 5}=k_{\mu\nu}$,   $h= g^{\mu\nu}k_{\nu\nu}$,    $K^{2}=k^{\mu\nu}k_{\mu\nu}$  and
\[
Q_{\mu\nu} = -k^{\rho}{}_{\mu }k_{\rho\nu }+ h
k_{\mu\nu}\!\!  +  \!\!\frac{1}{2}(K^{2}-h^{2})g_{\mu\nu},
\;\;\;Q^{\mu\nu}{}_{;\nu} =0,   \label{eq:Qij}
\]
The   confinement of ordinary   matter and  gauge  fields  corresponds  simply  to   $T_{\mu 5}=T_{55}=0$.  The  cosmological  constant  $\Lambda$   comes  after the  contracted Bianchi  identity  applied to the entire  left-hand  side  of  \rf{gtensor}.

\subsection{The  Cosmological Constant Problem}
Here  we refer  to    the original  cosmological constant problem
described in \cite{Weinberg}.
Using the semiclassical Einstein's  equations in general relativity  the   quantum  vacuum can be described  as   a  perfect fluid  with  state   equation  $p_v = -<\rho_v>=$constant  \cite{Zeldowich}:
\be
R_{\mu\nu} -\frac{1}{2}R g_{\mu\nu} +\Lambda g_{\mu\nu}=  8\pi  G T_{\mu\nu}^m  +
8\pi G <\rho_v>g_{\mu\nu}   \label{eq:semiclassical}
\ee
where  $T_{\mu\nu}^m$ stands  for the  classical  sources.
Comparing the  constant  terms in both sides of this  equation we  obtain  $ \Lambda/8\pi G  = <\rho_v>$, or  as it is commonly    stated, \emph{the  cosmological constant is the  vacuum  energy density}.
However,   current observations tell that
 $\Lambda/8\pi G  \approx  10^{-47} \;Gev^4$  (here,  $c=1$).
On the other hand, admitting that  quantum  field theory  holds  up
to the  Planck  scale,  the vacuum energy density  would  be
$<\rho_v>  \approx  (10^{19}Gev)^4 =  10^{76}\; Gev^4$. This  difference  cannot be resolved by  any  known theoretical   procedure in  quantum  field  theory. Even  supposing  that  quantum field theory holds  to  the Tev  scale or  less,  the  difference would be  still  too large  to  compensate.  This  difficulty has become  to known  as the   cosmological  constant problem.

Here  we    consider  that  the  above   difference is  not   only  numerical,  but   it is  mainly   conceptual,   resulting from  the superposition of   two  incompatible   ground  states  for the  gravitational  field  in general relativity:  The  flat Minkowski   ground  state was  chosen  to be the  reference  of  curvature,  but  the experimental  evidences  of   $\Lambda /8\pi G  \neq 0$  however  small,    point  to a  deSitter  ground  state,  which is  conceptually    incompatible  with   the Minkowski's choice. The
implications being that   particles and  fields, their  masses and  spins defined by the Casimir  operators of the deSitter  group  are   different  from  those  defined by the Poincaré group,  and they  coincide only  when  $\Lambda$  vanishes.

The above  numerical  and conceptual  conflicts    can  be resolved   with  the  Schlaefli   embedding   conjecture as  implemented  by Nash,    where the deSitter  and  Minkowski  space-times   may  coexist. Indeed,
 in  \rf{gtensor}   $\Lambda/8\pi G_* $  is  a gravitational  component   resulting from  the gravitational  equations  in the embedding space. However,  the vacuum energy  density $<\rho_v>$  is  a confined quantity in the space-time,  regardless  of the perturbations of  its metric. Finally,  the presence of the  extrinsic  curvature  $k_{\mu\nu}$ the  conserved quantity  $Q_{\mu\nu}$  of \rf{gtensor},  imply   that  those  constants   cannot  be   canceled without imposing  a  constraint on the  extrinsic  curvature,   which  is now  part  of the  gravitational dynamics in the embedding space.

\subsection{The Accelerated Expansion}

To  solve  the  system  \rf{gtensor}, \rf{gvector}  we   start  with  the last of  these  equations  to  determine  the general form of  the  extrinsic  curvature for  the  FLRW   metric.  We  find  that
$$
 k_{ij}=\frac{b}{a^2}g_{ij},\;\;i,j=1,2,3, \;\;\;\;
k_{44}=\frac{-1}{\dot{a}}\frac{d}{dt}\frac{b}{a}
$$
where   $b(t)\, (=k_{11})$  remains  an  arbitrary function as  a   consequence the  confinement  condition.
Replacing this  result   in \rf{gtensor} we obtain  the  modified
Friedman's  equation
\be
\left(\frac{\dot{a}}{a}\right)^2+\frac{\kappa}{a^2}=\frac{4}{3}\pi
G_*\rho+\frac{\Lambda}{3}+\frac{b^2}{a^4}  \label{eq:Friedmanx}
\ee
showing  that  the    extrinsic  curvature contributes to  the  accelerated   expansion of the universe.  To see  this, as  a first guess to  determine  $b(t)$   we have considered an analogy between   the    extrinsic  curvature and the  and x-matter (XCDM) phenomenological fluid  \cite{Turner}. Interpreting   $Q_{\mu\nu}$  as  a geometrical fluid
$$
Q_{\mu\nu}  =\frac{1}{8\pi G_*}[(p_x  -\rho_x)U_\mu U_\nu
 -\rho g_{\mu\nu}],\;\;  p_x  =\omega_x \rho_x
$$
and  comparing this  with the  geometric expression of
 $Q_{\mu\nu}$  for the  FLRW  universe,   a   simple  equation for  $b(t)$ is   obtained
\[
\frac{\dot{b}}{b} =\frac{1}{2}(1-3\omega_x)\frac{\dot{a}}{a}
\]
This   cannot be  integrated  because  we do not know  $\omega_x$.
However, assuming  in particular  that   $\omega_x  = constant =\omega_0$
we obtain  a  simple solution
$$
b(t)  =  b_0 (\frac{a}{a_0})^{\frac{1}{2}(1-3\omega_0)}
$$
Replacing in  \rf{Friedmanx}   we  obtain the  modified  Friedman's  equation
\be
\left(\frac{\dot{a}}{a}\right)^2+\frac{\kappa}{a^2}=\frac{4}{3}\pi
G_*\rho+\frac{\Lambda}{3}+  b_0^2\left(\frac{a}{a_0}\right)^{3(1 + \omega_0)}  \label{eq:Friedman}
\ee
where the     constant  $a_0$   corresponds  to  today's  radius of the universe.  In particular  using  $G=G_*$  the  constants $b_0$ and   $\omega_0$ can be   adjusted  to match   the  accelerated  expansion   of the  universe as  compared  with the  XCDM   phenomenological  model  \cite{GDEI}.

\subsection{The Dynamics  of the Extrinsic  Curvature}

With the  application  of  Nash's  theorem  to  cosmology,  the  extrinsic  curvature  assume the   fundamental  role of  promoting the  perturbations of the  gravitational  field.  Noting that it  is a  symmetric rank-2  tensor  field, it  represents  a spin-2 field in space-time,  whose  dynamics  must be  determined  independently of   the  fluid  model.

 In 1954  S. Gupta noted  that   the   Fierz-Pauli equation  for  a linear  massless spin-2  field in Minkowski  space-time
has  a  remarkable  resemblance  with  the  linear
approximation  of  Einstein's  equations  \cite{Gupta}.
He  found  that indeed  such  field must   satisfy   an equation    that   has  the  same  formal  structure  as  Einstein's equations,
defined  on  Minkowski's  space-time.

In order to derive  Gupta's  equation for $k_{\mu\nu}$  for   any curved  manifold,  we  use an  analogy  with  Riemannian's geometry,    keeping  in mind  that the geometry of the embedded space-time has been already defined  by the  metric  $g_{\mu\nu}$.

In the case   of  a  5-dimensional  embedding space,   we replace  $k_{\mu\nu}$  by  a  new   tensor  given   by
$$f_{\mu\nu} = \frac{2}{K}k_{\mu\nu}, \;\; \mbox{and}
\;\;f^{\mu\nu} = \frac{2}{K}k^{\mu\nu}$$
so  that  $f^{\mu\rho}f_{\rho\nu} =\delta^\mu_\nu  $.
Then,    the  analogous  to   the  Levi-Civita  connection is
$$
\Upsilon_{\mu\nu\rho}=\;\frac{1}{2}\left(
f_{\rho\nu,\mu}+ f_{\rho\mu,\nu}
-f_{\mu\nu,\rho}\right)\;\,\;\mbox{and}\,\;\;
\Upsilon_{\mu\nu}{}^{\sigma}=
f^{\sigma\rho}\;\Upsilon_{\mu\nu\rho}
$$
After, imposing   the equivalent to  "metricity  condition'   $f_{\mu\nu||\rho}=0$,    we may calculate   the analogous of  the  "Riemann curvature" tensor
$$ \mathcal{F}_{\nu\alpha\lambda\mu}=
\Upsilon_{\mu\lambda\nu,\alpha}-\Upsilon_{\mu\alpha\gamma,\lambda}+
\Upsilon_{\alpha\sigma\mu}\Upsilon_{\lambda\nu}^{\sigma}
-\Upsilon_{\lambda\sigma\mu}\Upsilon_{\alpha\nu}^{\sigma}
$$
and define   Ricci  and  scalar  tensors respectively by $\mathcal{F}_{\mu\nu} = f^{\alpha\beta}\mathcal{F}_{\mu\alpha\beta\nu}$,  $\mathcal{F}=f^{\mu\nu}\mathcal{F}_{\mu\nu}$.
With these definitions  we  arrive  at Gupta's  equations  for  the  f-field
\be
 \mathcal{F}_{\mu\nu}-\frac{1}{2}\mathcal{F} f_{\mu\nu}
=\;\alpha_f\tau_{\mu\nu}  \label{eq:gupta}
 \ee
where $\tau_{\mu\nu}$ acts  like  a  source field  for the extrinsic  curvature,  with       a  coupling  constant  $\alpha_f$.
A simple  example  is   given by the vacuum   equation  $ \mathcal{F}_{\mu\nu}=0$  which   reproduces  in particular  the
XCDM  model, but now  without   any  appeal  to  the  fluid  analogy  \cite{GDEII}.

\subsection{Structure Formation}

The  formation of  large  structures  in the  early  universe  has been  mostly   attributed  to   gravitational  perturbations   produced by   other  than  baryons   sources,   generally  referred  to  as the dark matter  component of the  universe. In the present case,  the   extrinsic  curvature   solution of  \rf{gupta}   should  have  an  observable  effect in  space-time,  independently of  the perturbations.  Therefore,  it is  possible  that  the theoretical  power  spectrum  obtained   from  \rf{Friedman} coincide  with the  observed  one.
In a   preliminary  analysis we  obtain a  power  spectrum  which is similar  to  the power  spectrum  from   the  cosmic  microwave  background  radiation  obtained  from   the  WMAP   experiment.
\begin{figure}[!h]\begin{center}
\includegraphics[width=8cm,height=5.5cm]{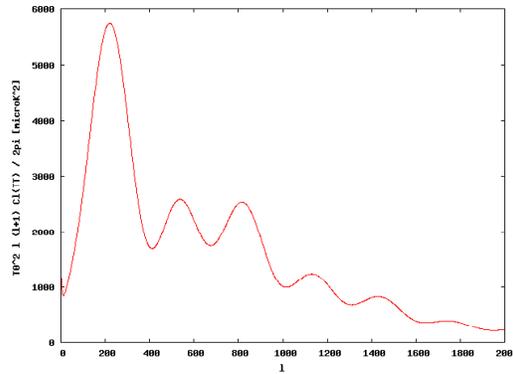}
\caption{\footnotesize The   NASA-CAMB theoretical power  spectrum  resulting from   \rf{Friedman}  calculated
for  $   -1 \le  \omega_0 \le -1/3$, Massive
Neutrinos=1, massless  neutrinos  =3.04.}\end{center}
\end{figure}\\
As  we  see,  the  third  peak  closely  matches  the  observed  one \cite{GDM2008}.

\newpage
\begin{center}
\textbf{Concluding Remarks}
\end{center}
The embedding  of a space-time  manifold   into  another  defined by the  Einstein-Hilbert principle  may  lead  to an interesting gravitational theory,  not only  because its  mathematical consistency provided by the Schlaefli  conjecture as  resolved by  Nash's theorem,  but mainly because it can meet  the demands  of modern  cosmology,  with the  minimum of  additional
assumptions.

\begin{center}
\textbf{Acknowledgements}
\end{center}
The author   wishes to thank   the  UFRN  and the organizers
of the  Second  Indo-Brazilian Workshop. In particular  he  wants  to express his thanks  to    Edmundo Monte,  Jailson Alcaniz and  Abraão Capistrano  for their valuable contributions  to  some results  on the  subject  of  this  talk.

\end{document}